\documentclass[12pt,a4paper,reqno]{article}
\usepackage{amssymb,amsmath}
\def\eqref#1{{\rm(\ref{#1})}}
\def\Id{\mathop{\mathrm{Id}}\nolimits}

\newtheorem{remark}{Remark}

\begin{document}
\title{Recursion operator for the IGSG equation}

\author{M. Marvan and M. Pobo\v{r}il 
\\\footnotesize Mathematical Institute in Opava, Silesian University in Opava, 
\\\footnotesize Na Rybn\'\i\v cku 1, 746 01 Opava, Czech Republic, 
\\\footnotesize M.Marvan@math.slu.cz, M.Poboril@math.slu.cz}

\date{April 11, 2006}
\maketitle

\abstract{In this paper we find the inverse and direct recursion operator for the intrinsic generalized sine-Gordon equation in any number~$n > 2$ of independent variables.
Among the flows generated by the direct operator we identify a higher-dimensional analogue of the pmKdV equation. 


{\bf MS Classification (2005). }
35Q53, 37K10, 53C42, 58J70.

}

\paragraph{Introduction. } General recursion operators as introduced by Olver~\cite{O} are pseudo\-differential operators $\sum_{i = -r}^s f_i D_x^i \circ h_i$ mapping symmetries to symmetries, thus capable of generating infinite series of them.
An important generalization by Guthrie~\cite{G} eliminates inherent problems of interpretation of $D_x^{-1}$.
Moreover, a link exists between Guthrie's definition and the theory of coverings (Krasil'shchik and Vinogradov~\cite{K-V 1,K-V 2}) which implies that main constituents of Guthrie recursion operators trivialize in higher dimensions unless the system in question is (maximally) overdetermined~\cite{M2}. 

The goal of this paper is to demonstrate that there do exist overdetermined systems in any dimension which admit a nontrivial recursion operator under Guthrie's definition, thus escaping the bi-local approach long established in multidimension \cite{B-L-P,F-S 1,F-S 2}. 

The system in question is the intrinsic generalized sine-Gordon equation (IGSGE) \cite{B-T}, describing metrics of $n$-dimensional Riemannian spaces $M$ of constant sectional curvature $K$ immersed into a $(2n - 1)$-dimensional Riemannian space $\bar M$ of constant sectional curvature $\bar K$. 
The subcase $K = 0$ of flat metrics on $M$ is usually referred to as the intrinsic generalized wave equation (IGWE).
These systems are known to be integrable in the sense of soliton theory. 
In particular, these equations posses a B\"acklund transformation~\cite{B-T} as well as soliton solutions~\cite{Ten,G-H}.
Closely related are the generalized sine-Gordon equation (GSGE) and generalized wave equation (GWE)~\cite{A,T1,T-T}, which describe the immersions themselves, as well as the Generalized Equation (GE) and the Intrinsic Generalized Equation (IGE)~\cite{C-T,B-F-T}, referring to pseudo-Riemannian metrics.
All these systems are integrable as well, with ever growing bibliography of results, e.g.,~\cite{C-A,F-P,T2,Zh}, to list just few.

Lie symmetries of IGSGE and IGWE were computed by Tenenblat and Winternitz~\cite{T-W} along with the corresponding invariant solutions, 
this result being extended to GE by Ferreira~\cite{F}. However, apparently no higher symmetries have been written yet.
We utilize the recursion operator found in this paper to generate a local flow of third order, which we call the generalized pmKdV equation, and conjecture it to be an integrable evolution system. 

Unless otherwise stated, indices range between $1$ and~$n$.

\paragraph{The equation.}
\label{IGSGE}
(\cite{B-T,Ten,T-W})
Let an $n$-dimensional Riemannian manifold $M$ of constant sectional curvature $K$ be immersed in an $(2n - 1)$-dimensional Riemannian manifold $ M$ of constant sectional curvature $ K$.
Then $M$ admits orthogonal coordinates $x^1,\dots,x^n$, $I = (v^1\,dx^1)^2 + \dots + (v^n\,dx^n)^2$, with the $n$ Lam\'e coefficients $v^1,\dots,v^n$ satisfying an algebraic constraint 
\begin{equation}
\label{sq}
\sum_i (v^i)^2 = 1, 
\end{equation}
and $n^3 - n^2 + n$ first-order differential equations
\begin{equation}
\label{Eq}
\begin{aligned}
&v^i_j = v^j h^{ji}, \qquad j \ne i, \\
&v^i_i = -\sum_{s \ne i} v^s h^{is}, \\
&h^{ij}_i = -h^{ji}_j - K v^i v^j - \sum_{s \ne i,j} h^{si} h^{sj},
  \qquad i > j, \\
&h^{ij}_j = -h^{ji}_i - \sum_{s \ne i,j} h^{is} h^{js},
  \qquad i < j, \\
&h^{ij}_k = h^{ik} h^{kj}, \qquad j \ne i \ne k \ne j. 
\end{aligned}
\end{equation}
Here $h^{ij}$, $i \ne j$, are $n^2 - n$ auxiliary functions (rotation coefficients).
The lower indices denote differentiation, hence $v^i_j$ stands for $\partial v^i/\partial x^j$ etc.

This system is referred to as IGSGE when $K \ne 0$ and IGWE when $K = 0$.
Our results, formulated for IGSGE, are easily transferable to IGWE.
In the sequel we apply the convention $h^{ii} = 0$ to remove the restrictions on summation indices.

To start with, we check the highly overdetermined system \eqref{Eq} for consistency. The system is orthonomic in the sense of the Riquier theory (see~\cite[\S~6]{Doug} for a compact summary) under a suitable ranking of derivatives. Namely, if derivatives $\partial^{r_1 + \dots + r_n}/(\partial x^1)^{r_1} \dots (\partial x^n)^{r_n}$ are ranked by their total order $r_1 + \dots + r_n$, with ties broken by the degree $r_n$ in $x^n$, with ties broken by the degree $r_{n-1}$ in $x^{n-1}$, etc., then \eqref{Eq} is (irrespectively of the ranking of the dependent variables $v^i, h^{ij}$) resolved with respect to the highest rank derivatives and hence so is its infinite prolongation resulting from differentiation of each equation of \eqref{Eq} with respect to an arbitrary combination of $x^i$'s.

Derivatives on the left-hand side of \eqref{Eq} and its infinite prolongation are called {\it principal\/} and become functions of the remaining -- {\it parametric} -- derivatives.
Now, the system~\eqref{Eq} is easily seen to be passive, meaning that all its integrability conditions (equality of cross derivatives) are identically satisfied and are so even without reference to the algebraic constraint~\eqref{sq}.
Moreover, one easily checks that the derivative $\sum_i v^i v^i_k = 0$ of Eq.~\eqref{sq} with respect to an arbitrary $x^k$ is satisfied by virtue of the first two equations of system~\eqref{Eq}. 
We then conclude that the whole system~\eqref{sq}, \eqref{Eq} is passive.

The constraint \eqref{sq} means that $(v^1,\dots,v^n)$ lies in $S^{n - 1}$, the unit sphere. Then $v^n$ can be expressed through $v^1,\dots,v^{n - 1}$ on each of the two hemispheres $v^n > 0$ and $v^n < 0$ of $S^{n - 1}$. 
Hence the complete list of parametric derivatives to be used below:
\begin{equation}
\begin{aligned}
\label{Q}
&v^1,\dots,v^{n-1}, \quad
\\ 
&h^{ij}, \quad i \ne j, \\ 
&h^{ij}_{\underbrace{\scriptstyle i \dots i}_r}, \quad i > j,\\ 
&h^{ij}_{\underbrace{\scriptstyle j \dots j}_r}, \quad i < j,
\end{aligned}
\end{equation} 
the number $r$ of lower $i$'s or $j$'s being arbitrary positive. 

In terms of geometric theory of systems of PDE \cite{K-V}, we have the following picture of the system~\eqref{sq},~\eqref{Eq}. 
Let $Y$ denote the product of the unit sphere $S^{n - 1}$, coordinatized by  $v^1,\dots,v^{n-1}$, and the arithmetic space $\mathbb R^{n(n - 1)}$, coordinatized by $h^{ij}$, $i \ne j$. Consider the trivial bundle $\pi : Y \times M \to M$ and its infinite prolongation $J^\infty \pi$.
Then system \eqref{Eq} and its infinite prolongation (consisting of all differential consequences) above determine a submanifold $\mathcal E$ in $J^\infty \pi$. The system being passive, differential consequences with identical principal derivatives on the left-hand side are equivalent. 
The parametric derivatives \eqref{Q} then serve as coordinates along the fibres of $\mathcal E \to M$.

Given a function $f \in C^\infty J^\infty \pi$, its restriction $f|_{\mathcal E}$ to $\mathcal E$ is computed if every principal derivative $f$ depends on is substituted by the right-hand side of the corresponding equation~\eqref{Eq} or its differential consequence until all principal derivatives disappear (which is guaranteed by the use of Riquier ranking). This leaves us with a function independent of derivatives of $v^i$. Then $v^n$ are replaced with their expression through the coordinates $v^1,\dots,v^{n-1}$. This is a workable procedure that leaves us with a function involving parametric derivatives~\eqref{Q} only.

The usual total derivatives with respect to $x^k$ admit restriction on $\mathcal E$, namely
$$
\begin{aligned}
&D_k = \frac\partial{\partial x^k}
 + \sum_{i = 1}^{n-1} v^i_k|_{\mathcal E} \frac\partial{\partial v^i}
 + \sum_{i < j} h^{ij}_k|_{\mathcal E} \frac\partial{\partial h^{ij}}
\\&\qquad
 + \sum_{r > 0} \sum_{i > j} h^{ij}_{\underbrace{\scriptstyle i \dots i}_r k}|_{\mathcal E}
   \frac\partial{\partial h^{ij}_{\underbrace{\scriptstyle i \dots i}_r}}
 + \sum_{r > 0} \sum_{i < j} h^{ij}_{\underbrace{\scriptstyle j \dots j}_r k}|_{\mathcal E}
   \frac\partial{\partial h^{ij}_{\underbrace{\scriptstyle j \dots j}_r}},
\end{aligned}
$$
where coefficients $v^i_k|_{\mathcal E}$, $h^{ij}_k|_{\mathcal E}$, etc. are determined from  system~\eqref{Eq} and its prolongations.

\paragraph{The zero-curvature representation} of system \eqref{Eq}, due to Beals and Tenenblat~\cite{B-T}, consists of sparse antisymmetric $2n \times 2n$ matrices ${\mathbf A}_{(k)}$, $k = 1,\dots,n$, given by
\begin{equation}
\begin{aligned}
\label{zcr}
&A_{(k)}^{ij} = \delta_k^j h^{ji} - \delta_k^i h^{ij}, \\
&A_{(k)}^{n+i,j} = -A_{(k)}^{i,n+j} = \left(\frac K{4z} - z\right) \delta_k^i \delta_k^j 
                - \frac K{2z} \delta_k^i v^j v^k, \\ 
&A_{(k)}^{n+i,n+j} = \delta_k^j h^{ij} - \delta_k^i h^{ji}.
\end{aligned}
\end{equation}
Here $z$ is the so-called spectral parameter.
The zero curvature condition 
${\mathbf A}_{(k) x^l} - {\mathbf A}_{(l) x^k}
 + [{\mathbf A}_{(k)},{\mathbf A}_{(l)}] = 0$,
which holds as a consequence of system \eqref{Eq}, is geometrically expressed as
\begin{equation}
\label{MC}
(D_l {\mathbf A}_{(k)} - D_k {\mathbf A}_{(l)}
 + [{\mathbf A}_{(k)},{\mathbf A}_{(l)}])|_{\mathcal E} = 0.
\end{equation}

\paragraph{Symmetries and recursion operators.}
As is well known (see, e.g., \cite{K-V}), a symmetry of PDE can be identified with a vertical vector field $Q = \sum_\iota Q^\iota\,\partial/\partial q^\iota$ on the corresponding manifold $\mathcal E \to M$. The field $Q$ is required to commute with the total derivatives $D_k$ restricted to $\mathcal E$. It turns out that symmetries of a system of equations $\{F^l = 0\}$ have to satisfy the {\it linearized system} $\{\ell_{F^l} = 0\}$, where 
$$
\ell_{F^l} = \sum_i \frac{\partial F^l}{\partial q^\iota} Q^\iota,
$$
with $q^\iota$ running through the coordinates on the manifold $\mathcal E$ (parametric derivatives).
Moreover, the condition of commutativity with $D_k$ implies that the coefficient $Q^\kappa$ corresponding to a derivative $q^\kappa = D_k q^\iota$ of another coordinate $q^\iota$ is expressible as $Q^\kappa = D_k Q^\iota$. Thus, the only  coefficients to be found are the $Q^j$ corresponding to variables $q^j$ of zeroth order ($v^i$ and $h^{ij}$ in our case). 

Hence, regarding our system~\eqref{sq}, \eqref{Eq}, symmetries are determined by functions $V^i$, $H^{ij}$ on the manifold $\mathcal E$ (i.e., depending on \eqref{Q}) that satisfy
\begin{equation} \label{lEq}
\begin{aligned}
&\sum_i v^i V^i = 0, \\
&V^i_j = v^j H^{ji} + h^{ji} V^j, \qquad j \ne i, \\
&V^i_i = -\sum_{s \ne i} (v^s H^{is} + h^{is} V^s), \\
&H^{ij}_i = -H^{ji}_j - K (v^i V^j + v^j V^i)
\\&\qquad\quad
 - \sum_{s \ne i,j} (h^{si} H^{sj} + h^{sj} H^{si}), \qquad i > j, \\
&H^{ij}_j = -H^{ji}_i - \sum_{s \ne i,j} (h^{is} H^{js} + h^{js} H^{is}),
 \qquad i < j, \\
&H^{ij}_k = h^{ik} H^{kj} + h^{kj} H^{ik}, \qquad j \ne i \ne k \ne j, 
\end{aligned}
\end{equation}
where $V^i_j = D_j V^i$ etc. 

Following Guthrie~\cite{G} and our earlier observation~\cite{M2}, we interpret recursion operators as B\"acklund autotransformations for the linearized system~\eqref{lEq}.

As a rule~\cite{M-S,B-M}, there exists a recursion operator that can be written in terms of an auxiliary system of equations closely related to the zero-curvature representation, namely
\begin{equation}
\label{W}
{\mathbf W}_{x^k}
 = [{\mathbf A}_{(k)}, {\mathbf W}] + \ell_{{\mathbf A}_{(k)}} {\mathbf W},
\end{equation}
with $\ell_{{\mathbf A}_{(k)}}$ computed componentwise.
In our case, ${\mathbf W}$ is an antisymmetric $2n \times 2n$ matrix, akin to matrices ${\mathbf A}_{(k)}$.
Compatibility of system \eqref{W} follows from the zero curvature condition \eqref{MC}.

Now, it is easily checked that if $V^i,H^{ij}$ are symmetries and ${\mathbf W}$ satisfies \eqref{W}, then $V^{\prime i}, H^{\prime ij}$ given by
\begin{equation}
\label{iRO}
\begin{aligned}
&V^{\prime i} = 2 z V^i - 2 z \sum_{s \ne i} v^s W^{is}, \\
&H^{\prime ij} = K v^j \sum_{s = 1}^n v^s W^{s,n+i} - (\tfrac12 K + 2 z^2) W^{j,n+i}
\end{aligned}
\end{equation}
are symmetries as well.
Thus, formulas \eqref{W} and \eqref{iRO} determine a family of recursion operators $\mathcal R_z$ for the IGSG equation.
When applied to a (local) symmetry of IGSG, $\mathcal R_z$ produces as much as $\frac12 n(n - 1)$ symmetries, all of them highly nonlocal as a rule. 
In such cases the conventional recursion operator is to be found among inverses of $\mathcal R_z - \lambda \Id$ for suitable $\lambda$ and appropriate choice of the spectral parameter $z$.

\paragraph{The conventional recursion operator} 
turns out to coincide with $\mathcal R_z^{-1}$ whenever $K \ne \pm4 z^2$. 
Inversion of $\mathcal R_z$ (for arbitrary $n$) uses~\eqref{W}, \eqref{iRO}, the linearized equation~\eqref{lEq} and the same linearized equation~\eqref{lEq} written for $V^{\prime i}, H^{\prime ij}$ (referred to as (\ref{lEq}$'$)) to express $V^i,H^{ij}$ in terms of $V^{\prime i},H^{\prime ij}$ while eliminating as many $W^{ij}$ as possible. 

To start with, we find
$$
V^i = \frac1{2z} V^{\prime i} + \sum_{s \ne i} v^s W^{is},
$$
from the first equation of \eqref{iRO} and
$$
\begin{aligned}
&H^{ij} = -D_i W^{ij} + \sum_s h^{is} W^{js}
\\&\qquad + \left(\frac{K}{4 z} - \frac{K}{2 z}(v^i)^2 - z\right) W^{j,n+i}
 + \frac{K}{2 z} v^i v^j W^{i,n+i}
\end{aligned}
$$
from the upper left off-diagonal part of \eqref{W}.
For all $i \ne j$, we then have
$$
\begin{aligned}
&W^{i,n+j} = -2 \frac
{H^{\prime ji}}
{K + 4 z^2}
 + \frac
{4 K v^i \sum_s v^s H^{\prime js}}
{(K + 4 z^2) (K - 4 z^2 - 2 K (v^j)^2)} 
\\&\qquad\quad - \frac
{2 K v^i v^j W^{j,n+j}}
{K - 4 z^2 - 2 K (v^j)^2} 
\end{aligned}
$$
from the second equation of \eqref{iRO}, assuming that $K \ne -4 z^2$. 
Similarly, all remaining components $W^{ij}$ and $W^{n+i,n+j}$ can be expressed algebraically in terms of $W^{i,n+i}$, assuming additionally that $K \ne 4 z^2$.
Omitting details of tedious computation, we finally obtain $W^{i,n+i}$ in terms of ``potentials'' $Q^i$, which can be characterized as follows.
Let
$$
\begin{aligned}
&q^i_i = K (v^i)^2 + \sum_s ((h^{si})^2 + (h^{is})^2), \\
&q^i_k = -2 h^{ik} h^{ki} \qquad \text{for $i \ne k$}.
\end{aligned}
$$    
Then $D_l q^i_k = D_k q^i_l$ for all $k \ne l$ on $\mathcal E$, meaning that horizontal forms $\xi^i = q^i_k\,dx^k$ determine an abelian covering over \eqref{Eq} (see~\cite{Kh}). 
Linearized forms $\Xi^i = Q^i_k\,dx^k$ with
\begin{equation}
\begin{aligned}
\label{RO1}
Q^i_i = 2 K v^i V^{\prime i}
 + 2 \sum_s (h^{si} H^{\prime si} + h^{is} H^{\prime is}), \\
Q^i_k = -2 h^{ik} H^{\prime ki} - 2 h^{ki} H^{\prime ik}
 \qquad \text{for $i \ne k$}
\end{aligned}
\end{equation}
then satisfy $D_l Q^i_k = D_k Q^i_l$ 
on (\ref{lEq}$'$), hence determine an abelian covering over the linearized equation (\ref{lEq}$'$). 
The required potentials $Q^i$ are just naturally chosen nonlocal variables of this covering and may be thought of as independent quantities satisfying
\begin{equation}
\label{RO2}
D_k Q^i = Q^i_k. 
\end{equation}
In terms of potentials $Q^i$, the final result of inversion of $\mathcal R_z$ is
\begin{equation}
\label{RO3}
\begin{aligned}
&V^i = 2 \sum_s v^s D_i H^{\prime is}
     + 2 \sum_t h^{ti} \sum_s v^s H^{\prime ts}
     - 2 \sum_t h^{ti} \sum_s v^i H^{\prime ti}
\\&\quad     - 2 K (v^i)^2 V^{\prime i}
     + \sum_s v^s (h^{is} Q^i - h^{si} Q^s),
\\
&H^{ij} = -2 D_i^2 H^{\prime ij}
     - 2 \sum_s h^{is} D_s H^{\prime sj} 
     - 2 \Bigl(K (v^i)^2 + \sum_s (h^{si})^2\Bigr) H^{\prime ij} \\&\quad
     - 2 \sum_s \Bigl(h^{ij} h^{is} H^{\prime is} 
       + 2 h^{ij} h^{si} H^{\prime si} 
       + \Bigl(h^{si}_i + \sum_{t \ne j} h^{it} h^{st}\Bigr) H^{\prime sj}\Bigr) \\&\quad
     - 2 K h^{ij} (2 v^i V^{\prime i} - v^j V^{\prime j}) \\&\quad
     - h^{ij}_i Q^i    
     + \Bigl(h^{ji}_i + \sum_s h^{is} h^{js}\Bigr) Q^j    
     - \sum_s h^{is} h^{sj} Q^s
\end{aligned}
\end{equation}
(terms containing $z$ divided out).
Equations (\ref{lEq}$'$), \eqref{RO1}--\eqref{RO3} imply (\ref{lEq}).
That is, \eqref{RO1}--\eqref{RO3} determine a recursion operator, henceforth denoted $\mathcal L$, for the IGSG equation.

\begin{remark} \rm
\label{rem}
The total derivatives $D_k$ in \eqref{RO2} are those of the covering equation (they commute by virtue of $D_l Q^i_k = D_k Q^i_l$) and hence the equation \eqref{RO2} is somewhat formal.
Only when $\mathcal L$ is to be applied to a seed symmetry $\sigma$ with components $V^{\prime i}, H^{\prime ij}$, then the $Q^i_k$, as well as $\Xi^i = Q^i_k\,dx^k$, induce objects on $\mathcal E$ (or an appropriate covering thereof if the seed is a nonlocal symmetry). It happens rather often that the induced forms $\Xi^i$ are exact on $\mathcal E$, hence the potentials $Q^i$ exist as functions on $\mathcal E$ and the resulting symmetry $\mathcal L(\sigma)$ is local. If $\Xi^i$ are not exact on $\mathcal E$, then they induce an abelian covering $\tilde{\mathcal E}$ over $\mathcal E$ and $\mathcal L(\sigma)$ is then a nonlocal symmetry with shadow defined on $\tilde{\mathcal E}$~\cite{K-V 2,M2}. (This usually happens with $\mathcal R_z(\sigma)$ and explains why operators $\mathcal R_z$ fail to generate local symmetries.)
\end{remark}

\paragraph{Other cases.}
Routine computation shows that $\mathcal R_z - \lambda \Id$ is invertible for all $\lambda \ne 0$, but symmetry-generating properties of $(\mathcal R_z - \lambda \Id)^{-1}$ are no better than those of $\mathcal R_z$ itself.
Therefore we continue with $\lambda = 0$.

In two singular cases, when $K = \pm 4 z^2$, the B\"acklund autotransformation $\mathcal R_z$ admits a reduction of the set of nonlocal variables (see \cite{M3} for generalities).  
If $K = -4 z^2$, then $\mathcal R_z$ can be written in terms of $2n$ nonlocal variables $p^i = \sum_{j=1}^n v^j W^{ij}$, $i = 1,\dots,2n$ as
$$
V^{\prime i} = 2 z (V^i - p^i), \qquad
H^{\prime ij} = 4 z^2 v^j p^{n+i},
$$
with $p^i$ being subject to another determining linear system 
$p^i_{x^k} = B^{ij}_{(k)} p^j + B^{\prime i}_{(k)}$,
whose exact form is not important here.
It follows immediately from the second equation that $H^{\prime i1}/v^1 = \dots = H^{\prime n1}/v^n$; therefore the image of $\mathcal R_z$ is not the full $\mathcal E$, hence $\mathcal R_z$ is noninvertible. 

When $K = 4 z^2$, a maximally reduced set of nonlocal variables is $p^i = \sum_{j=1}^n v^j W^{ij}$ with $i = 1,\dots,n$ and $q^{ij} = v^j W^{i,n+j} - v^i W^{j,n+j}$ with $i,j = 1,\dots,n$.
We can write $\mathcal R_z$ in terms of $p^i$, $q^{ij}$ as
$$
V^{\prime i} = 2 z (V^i - p^i), \qquad
H^{\prime ij} = 4 z^2 \Bigl(\sum_s \frac{v^s v^j}{v^i} q^{si}
 - \frac{q^{ji}}{v^i}\Bigr),
$$
omitting the rather cumbersome linear system for $p^i$, $q^{ij}$. Similarly as in the preceding case, detailed investigation reveals that the operator is not invertible.

\paragraph{Hierarchy of symmetries} generated by the operator $\mathcal L$ is expected to contain infinitely many commuting symmetries, as is usual with other integrable equations. However, since proof of this fact is still pending, we present just first members of the hierarchy.

As explained in Remark \ref{rem} above, application of the Guthrie operator $\mathcal L$ to a seed symmetry $S$ involves solution of the equation \eqref{RO2} with coefficients given by \eqref{RO1} with $V^{\prime i},H^{\prime ij}$ replaced by the components of the seed symmetry.

Applying $\mathcal L$ to the zero symmetry, we infer from \eqref{RO1}, \eqref{RO2} that $Q^i$ are constants, henceforth denoted by $-c^i$.
Then from \eqref{RO3} we obtain an $n$-dimensional algebra of first-order symmetries
$$ 
\begin{aligned}
&V^i = \sum_s (v^s h^{si} c^s - v^s h^{is} c^i),
\\
&H^{ij} = h^{ij}_i c^i + h^{ij}_j c^j + \sum_s h^{is} h^{sj} c^s,
\end{aligned}
$$
parametrized by $c^1,\dots,c^n$.
By virtue of \eqref{Eq} these can be identified with the obvious $x$-translations
$V^i = \sum_s v^i_s c^s$,
$H^{ij} = \sum_s h^{is}_s c^s$.
These were proved by Tenenblat and Winternitz~\cite{T-W} to exhaust the algebra of Lie symmetries when $K \ne 0$.

Application of $\mathcal L$ to the $x$-translations requires solution of system \eqref{RO2} under substitutions 
$V^{\prime i} = \sum_s (v^s h^{si} c^s - v^s h^{is} c^i)$,
$H^{\prime ij} = h^{ij}_i c^i + h^{ij}_j c^j + \sum_s h^{is} h^{sj} c^s$.
With $Q^i_k$ determined from \eqref{RO1}, the potentials $Q^i$ are just local functions
$$
Q^i = \Bigl(K (v^i)^2 + \sum_s (h^{is})^2 + \sum_s (h^{si})^2\Bigr) c^i
 - 2 \sum_s h^{is} h^{si} c^s.
$$
The resulting symmetry then has the following components
$$
\begin{aligned}
&V^i =
    - 2 \sum_s h^{is}_{ii} v^s c^i \\&\quad
    + \Bigl(\sum_s v^s h^{is}\Bigr)
          \Bigl(3 K (v^i)^2 + \sum_s (h^{is})^2 + 3 \sum_s (h^{si})^2\Bigr)
          c^i \\&\quad
    + 2 \sum_s h^{si}_{ss} v^s c^s 
    + 2 \sum_s \sum_t v^t (h^{si}_s h^{st} - h^{st}_s h^{si}) c^s \\&\quad
     + \sum_s v^s h^{si}
          \Bigl(3 K (v^s)^2 + \sum_t(h^{st})^2 + 3 \sum_t (h^{ts})^2\Bigr)
          c^s
\end{aligned}
$$
and
$$
\begin{aligned}
&H^{ij} =
   2 h^{ij}_{iii} c^i 
   + 6 h^{ij} \sum_s h^{si} h^{si}_i c^i 
   - 6 K v^i h^{ij} \sum_s v^s h^{is} c^i \\&\quad
   + h^{ij}_i \Bigl(3 K (v^i)^2
      + 3 \sum_s (h^{is})^2 + \sum_s (h^{si})^2\Bigr) c^i \\&\quad
   + 2 h^{ij}_{jjj} c^j 
   + 6 h^{ij} \sum_s h^{js} h^{js}_j c^j \\&\quad
   + h^{ij}_j \Bigl(3 K (v^j)^2
      + 3 \sum_s (h^{js})^2 + \sum_s (h^{sj})^2\Bigr) c^j \\&\quad
   + 2 \sum_s (h^{is}_{ss} h^{sj} - h^{is}_s h^{sj}_s
        + h^{sj}_{ss} h^{is} )c^s \\&\quad
   + 3 \sum_s h^{is} h^{sj} \Bigl(K (v^s)^2
      + \sum_t (h^{st})^2 + \sum_t (h^{ts})^2\Bigr) c^s
\end{aligned}
$$
(integration constants of~\eqref{RO2} are now omitted; they would just generate additional $x$-translations).

Consider the basis of symmetries \smash{$V^i_k$, $H^{ij}_k$} obtained by substitution $c_i = \delta_{ik}$. Interestingly enough, expressions \smash{$V^i_k$, $H^{ij}_k$} contain only derivatives with respect to $x^k$ for each $k$. In particular, 
$\{v^i_\tau = V^i_k, h^{ij}_\tau = H^{ij}_k\}$ is a system of evolution equations in $\tau$ and $x^k$, henceforth referred to as a {\it generalized pmKdV equation}.
The motivation is that for $n = 2$, eq.~\eqref{Eq} becomes the sine--Gordon equation $w_{y^1,y^2} = -K \sin w$ under identification $v^1 = \cos \frac12 w$, 
$v^2 = \sin \frac12 w$, $h^{12} = \frac12 w_{x^1}$, $h^{21} = -\frac12 w_{x^2}$, $y^1 = x^1 + x^2$, $y^2 = x^1 - x^2$~\cite{Ten}.
The same identification makes $V^i_1$ into the pmKdV equation
$w_t = w_{xxx} + \frac 12 w_x^3$ (with additional removable term $3 K w_x$).
The pmKdV equation is known to coincide with the first flow of the hierarchy associated with the sine--Gordon equation~\cite{Ku}.
We conjecture the generalized pmKdV equations and their higher analogues to be integrable.  

Hopes are that the hierarchy of generalized pmKdV equations in two independent variables possesses a bi-Hamiltonian formulation and a classical recursion operator, and the locality of this hierarchy can be proved using the results of Sergyeyev~\cite{Se}.
These and other interesting questions will be addressed elsewhere.

\paragraph{Acknowledgements.}
Both authors gratefully acknowledge support from GA\v{C}R under project No.~201/04/0538 and from M\v{S}MT under grant MSM 4781305904.


\begin{thebibliography}{99}


\bibitem{A}
Yu. Aminov, On immersions of regions of the $n$-dimensional Lobachevsky space into $(2n - 1)$-dimensional Euclidean space, {\it Dokl. Akad. Nauk SSSR} {\bf 236} (1977) 521--524 (in Russian); English transl. {it Sov. Math. Dokl.} {\bf 18} (1977) 1210-1213.

\bibitem{B-M}
H. Baran and M. Marvan, A conjecture concerning nonlocal terms of recursion operators, this issue.

\bibitem{B-F-T}
J.L.~Barbosa, W.~Ferreira and K.~Tenenblat,
Submanifolds of constant sectional curvature in pseudo-Riemannian manifolds,
{\it Ann. Global. Anal. Geom.} {\bf 14} (1996) 381--401.

\bibitem{B-T} %
R. Beals and K. Tenenblat, An intrinsic generalization for the wave and sine--Gordon equations, in: B. Lawson et al., eds., {\it Differential Geometry} Pitman Monographs 52 (Longman, 1991) 25--46. 


\bibitem{B-L-P}
M.~Boiti, J.J.-P.~L\'eon and F.~Pempinelli,
Canonical and noncanonical recursion operators in multidimensions,
{\it Stud. Appl. Math.} {\bf 78} (1988) 1--19.

\bibitem{C-T}
P.T.~Campos and K.~Tenenblat,
B\"acklund transformations for a class of systems of differential equations,
{\it Geom. Funct. Anal.} {\bf 4} (1994) (3) 270--287.

\bibitem{C-A}
J.~Cie\'sli\'nski and Yu.A.~Aminov,
A geometric interpretation of the spectral problem for the generalized sine-Gordon system,
{\it J. Phys. A: Math. Gen.} {\bf 34} (2001) L153--L159.

\bibitem{Doug}
J.~Douglas,
Solution of the inverse problem of the calculus of variations,
{\it Trans. Amer. Math. Soc.} {\bf 50} (1941) 71--128.

\bibitem{F}
W.~Ferreira,
On metrics of constant sectional curvature,
{\it Matem. Contempor\^anea} {\bf 9} (1995) 91--110.

\bibitem{F-P}
D. Ferus, and F. Pedit,
Isometric immersions of space forms and soliton theory. 
{\it Math. Ann.} {\bf 305} (1996) 329--342.

\bibitem{F-S 1}
A.S.~Fokas and P.M.~Santini,
Recursion operators and bi-Hamiltonian structure in multidimensions. I,
{\it Commun. Math. Phys.} {\bf 115} (1988) 375--419.

\bibitem{F-S 2}
A.S.~Fokas and P.M.~Santini,
Recursion operators and bi-Hamiltonian structure in multidimensions. II,
{\it Commun. Math. Phys.} {\bf 116} (1988) 449--474.

\bibitem{G-H}
C.H.~Gu and H.S.~Hu,
Explicit solutions to the intrinsic generalization for the wave and sine-Gordon equations,
{\it Lett. Math. Phys.} {\bf 29} (1993) 1--11.

\bibitem{G}
G.A.~Guthrie,
Recursion operators and non-local symmetries,
{\it Proc. R. Soc. London A} {\bf 446} (1994) 107--114.


\bibitem{Kh}
N.G.~Khor'kova,
Zakony sokhraneniya i nelokal'nye simmetrii,
{\it Matem. Zametki} {\bf 44} (1988) (1) 134--144.

\bibitem{K-V} 
I.S. Krasil'shchik and A.M. Vinogradov, eds., {\it
  Symmetries and Conservation Laws for Differential Equations of
  Mathematical Physics}, Translations of Mathematical Monographs.
  Vol.182, American Mathematical Society, Providence RI, 1999.

\bibitem{K-V 1}
I.S.~Krasilshchik and A.M.~Vinogradov,
Nonlocal symmetries and the theory of coverings: An addendum to A.M. Vinogradov's `Local symmetries and conservation laws',
{\it Acta Appl. Math.} {\bf 2} (1984) 79--96.

\bibitem{K-V 2}
I.S.~Krasil'shchik and A.M.~Vinogradov,
Nonlocal trends in the geometry of differential equations: symmetries, conservation laws, and B\"acklund transformations,
{\it Acta Appl. Math.} {\bf 15} (1989) 161--209.

\bibitem{Ku}
S. Kumei, 
Invariance transformations, invariance group transformations, and invariance groups of the sine-Gordon equations, {\it J. Math. Phys.} {\bf 16} (1975) 2461--2468.

\bibitem{M2}
M.~Marvan,
Another look on recursion operators, 
in: {\it Differential Geometry and Applications},
Proc. Conf. Brno, 1995 (Masaryk University, Brno, 1987) 393--402;
ELibEMS http://www.emis.de/proceedings.

\bibitem{M3}
M. Marvan, Some local properties of B\"acklund relations, {\it Acta Appl. Math.} {\bf 54} (1998) 1Ð25.

\bibitem{M4}
M. Marvan, 
Scalar second-order evolution equations possessing an irreducible 
$sl_2$-valued zero-curvature representation, 
{\it J. Phys. A: Math. Gen.} {\bf 35} (2002) 9431--9439. 

\bibitem{M-S}
M. Marvan and A. Sergyeyev, 
Recursion operator for the Nizhnik--Veselov--Novikov equation, 
{\it J. Phys. A: Math. Gen.} {\bf 36} (2003) L87--L92. 

\bibitem{O}
P.J.~Olver,
Evolution equations possessing infinitely many symmetries,
{\it J. Math. Phys.} {\bf 18} (1977) 1212--1215.


\bibitem{Se}
A. Sergyeyev, Why nonlocal recursion operators produce local symmetries: new results and applications, {\it J. Phys. A: Math. Gen.} {\bf 38} (2005) 3397-3407.

\bibitem{Ten}
K. Tenenblat, A note on solutions for the intrinsic generalized wave and sine--Gordon equation, {\it J. Math. Anal. Appl.} {\bf 166} (1992) 288--301.

\bibitem{T-W}
K. Tenenblat and P. Winternitz, On the symmetry groups of the intrinsic generalized wave and sine--Gordon equations, {\it J. Math. Phys} {\bf 34} (1993) 3527--3542.

\bibitem{T-T}
K. Tenenblat and Chuu Lian Terng, A higher dimension generalization of the sine-Gordon equation and its B\"acklund transformation,
{\it Bull. Amer. Math. Soc. (N.S.)} {\bf 1} (1979) 589--593.

\bibitem{T1}
C.L.~Terng,  
A higher dimension generalization of the sine-Gordon equation and its soliton theory. 
{\it Ann. of Math.} {\bf 111} (1980) 491--510.

\bibitem{T2}
C.L.~Terng,
Soliton equations and differential geometry,
{\it J. Diff. Geom.} {\bf 45} (1997) 407--445.


\bibitem{Z-K}
V.E.~Zakharov and B.G.~Konopelchenko,
On the theory of recursion operator,
{\it Commun. Math. Phys.} {\bf 94} (1984) 483--509.

\bibitem{Zh}
Z.X.~Zhou,
Darboux transformations for the twisted $so(p,q)$ system and local isometric immersions of space forms,
{\it Inverse Problems} {\bf 14} (1998) 1353--1370.


\end{thebibliography}
\end{document}